\documentclass[%
 prd,
 reprint,
superscriptaddress,
preprintnumbers,
nofootinbib,
 amsmath,amssymb,
 aps,
]{revtex4-1}
\pdfoutput=1
\usepackage{hyperref}
\usepackage[textwidth=1.25cm,textsize=footnotesize,
]{todonotes}

\usepackage[utf8]{inputenc}

\newcommand{\eV}{\text{\,eV}}

\newcommand{\MeV}{\text{\,MeV}}

\newcommand{\be}{\begin{equation}}
\newcommand{\ee}{\end{equation}}
\newcommand{\ba}{\begin{eqnarray}}
\newcommand{\ea}{\end{eqnarray}}

\renewcommand{\l}{\left(}
\renewcommand{\r}{\right)}

\begin{document}

\preprint{INR-TH/2021-017}

\title{
  BEST Impact on Sterile Neutrino Hypothesis\\
%  or
%  \\
%Results from the Baksan Experiment on Sterile Transitions
  %  (BEST): detailed analysis
}  
\author{Vladislav Barinov}
\email{barinov.vvl@gmail.com} 
\affiliation{Institute for Nuclear Research of the Russian Academy of Sciences,
  Moscow 117312, Russia}
\affiliation{Physics Department, Moscow State University, 
Vorobievy Gory, Moscow 119991, Russia}

\author{Dmitry Gorbunov}
\email{gorby@ms2.inr.ac.ru}
\affiliation{Institute for Nuclear Research of the Russian Academy of Sciences,
  Moscow 117312, Russia}
\affiliation{Moscow Institute of Physics and Technology, 
  Dolgoprudny 141700, Russia}

%\date{}

\begin{abstract}
Recently the Baksan Experiment on Sterile Transitions (BEST) has
presented results\,\cite{Barinov:2021asz} confirming the {\it gallium
anomaly}---a lack of electron neutrinos $\nu_e$ at calibrations of SAGE\,\cite{Abdurashitov:1998ne,Abdurashitov:2005tb} and
GALLEX\,\cite{Kaether:2010ag}---at the statistical significance
exceeding 5\,$\sigma$. This result is 
consistent with explanation of the gallium anomaly as electron
neutrino 
oscillations into sterile neutrino, $\nu_s$. Within this explanation
the BEST experiment itself provides the
strongest evidence for the sterile neutrino among all the previous 
anomalous results in the neutrino sector. 
We combine the results of gallium experiments with 
searches for sterile neutrinos in reactor antineutrino experiments (assuming
CPT-conservation in the $3+1$ neutrino sector). 
While the ``gallium'' best-fit point in the model parameter space
(sterile neutrino mass squared $m_{\nu_s}^2\approx 1.25$\,eV$^2$,
sterile-electron neutrino mixing $\sin^22\theta\approx 0.34$) is
excluded by these searches, a part of the  BEST-favored 2\,$\sigma$
region with $m^2_{\nu_s}>5$\,eV$^2$ is consistent with all of
them. Remarkably, the regions advertised by anomalous
results of the NEUTRINO-4 experiment\,\cite{NEUTRINO-4:2018huq}
overlap with those of the BEST
experiment: the best-fit point of the joint analysis is 
$\sin^22\theta\approx 0.38$, $m_{\nu_s}^2\approx7.3$\,eV$^2$, the
favored region  will be
explored by the KATRIN experiment\,\cite{KATRIN:2020dpx}.   
The sterile neutrino explanation
of the BEST results would suggest not only the extension of the
Standard Model of particle physics, but also either serious modifications of
the Standard Cosmological Model and Solar Model, or a specific
modification of the sterile sector needed to suppress the sterile neutrino
production in the early Universe and in the Sun. 
\end{abstract}

\maketitle

%%%%%%%%%%%%%%%%%%%%%%%%%%%%%%%%%%%%%%%%%%%%%%%%%%%%%%%%%%%%%%%%%
%%%%%%%%%%%%%%%%%%%%%%%%%%%%%%%%%%%%%%%%%%%%%%%%%%%%%%%%%%%%%%%%%
\section{Introduction}
%\label{sec:Intro}

%{\bf 1.}
Neutrino oscillations provide us with the only direct evidence for
incompleteness of the Standard Model of particle physics (SM), which
fails to explain them. While the fundamental physics responsible for
the oscillations is still elusive, results of the majority of
neutrino experiments can be phenomenologically described by
introducing neutrino mass matrix. The joint analysis (see
e.g. Ref.\,\cite{Esteban:2020cvm} for a recent one) of the oscillation data 
determines three mixing angles between the mass and flavor neutrino
eigenstates and two squared-mass differences.

However, results of several experiments
\cite{LSND:1996ubh,Abdurashitov:1998ne,Abdurashitov:2005tb,Kaether:2010ag,
  Mueller:2011nm,Huber:2011wv,MiniBooNE:2018esg,NEUTRINO-4:2018huq,IceCube:2020phf}
apparently cannot be described in this
way\,\cite{Kopp:2013vaa,Dentler:2018sju},
and within the oscillation picture ask for neutrino states
additional to those three of the SM. These results are commonly known 
as {\it neutrino anomalies} and the new states are known as {\it sterile
  neutrinos}, see e.g.\,\cite{Giunti:2019aiy}, since they mix with the SM (or {\it active}) neutrinos
but do not participate in the SM gauge interactions.

All the neutrino anomalies must be thoroughly
investigated\,\cite{Abazajian:2012ys,Boser:2019rta}. Indeed,
any unrecognized systematics constrain the application of neutrino
as, e.g. \cite{Sramek:2012nk,Adelberger:2010qa}
a tool in studying the inner structures of the Earth and Sun. On the other side,
the possible presence of sterile neutrinos in nature will be the first
example of the non-SM physics for which we have been looking for decades. Moreover,
the sterile neutrinos with mixing parameters required to explain the
anomalies\,\cite{Boser:2019rta}
are expected to be produced in the early Universe and be in
equilibrium with the primordial plasma before the epoch of Big Bang
Nucleosynthesis. They would change the standard predictions for the
primordial abundance of the light chemical elements as well as the
subsequent late-time history of our Universe, including the
recombination and formation of the large-scale cosmic
structures. Thus, in the present concordance $\Lambda$CDM
cosmological model these sterile neutrinos are
forbidden\,\cite{Planck:2018vyg}. The
situation was quite the opposite about ten years ago, when
both state-of-the-art analyses of the Big Bang Nucleosynthesis\,\cite{Izotov_2010} and
cosmic microwave background anisotropy\,\cite{WMAP:2010qai} favored the sterile
neutrino hypothesis. Anyway, with more complicated sterile neutrino sector
they can be safe for cosmology, see e.g.\,\cite{Hannestad:2013ana,Dasgupta:2013zpn,Chu:2015ipa}.
Moreover, presently we have several tensions in consistent
description of the cosmological data, the most serious are so-called the Hubble
tension and $\sigma_8$ tension. So far we have no natural
straightforward resolutions, but some suggestions in literature, see
e.g.\,\cite{Zhao:2017urm,Archidiacono:2020yey}  
involve sterile neutrinos as one of the necessary ingredients.    

%{\bf 2.}
\section{Results of the BEST experiment}
Therefore, both particle physics and cosmology would benefit from
settling this issue with neutrino anomalies, and there are various
projects and ongoing experiments dedicated to clarification of 
particular anomalous results. One of these experiments is Baksan
Experiment on Sterile Transitions (BEST) aimed at exploring the {\it
  gallium anomaly}\,\cite{Laveder:2007zz}, 
which is the lack of electron neutrino events from
compact artificial chromium and argon sources observed by four
(in total) calibrations of SAGE\,\cite{Abdurashitov:1998ne,Abdurashitov:2005tb} and GALLEX\,\cite{Kaether:2010ag} solar neutrino
experiments. BEST was placed in the Baksan Neutrino Observatory of INR RAS
and operated in 2019. Closely following the scheme of the original
SAGE calibration its geometry has been specifically designed to achieve the
highest sensitivity to the sterile neutrino model with parameters
tuned to explain the gallium anomaly via $\nu_e\leftrightarrow\nu_s$
oscillations.

\begin{figure}[!htb]
  \centerline{
\includegraphics[width=\columnwidth]{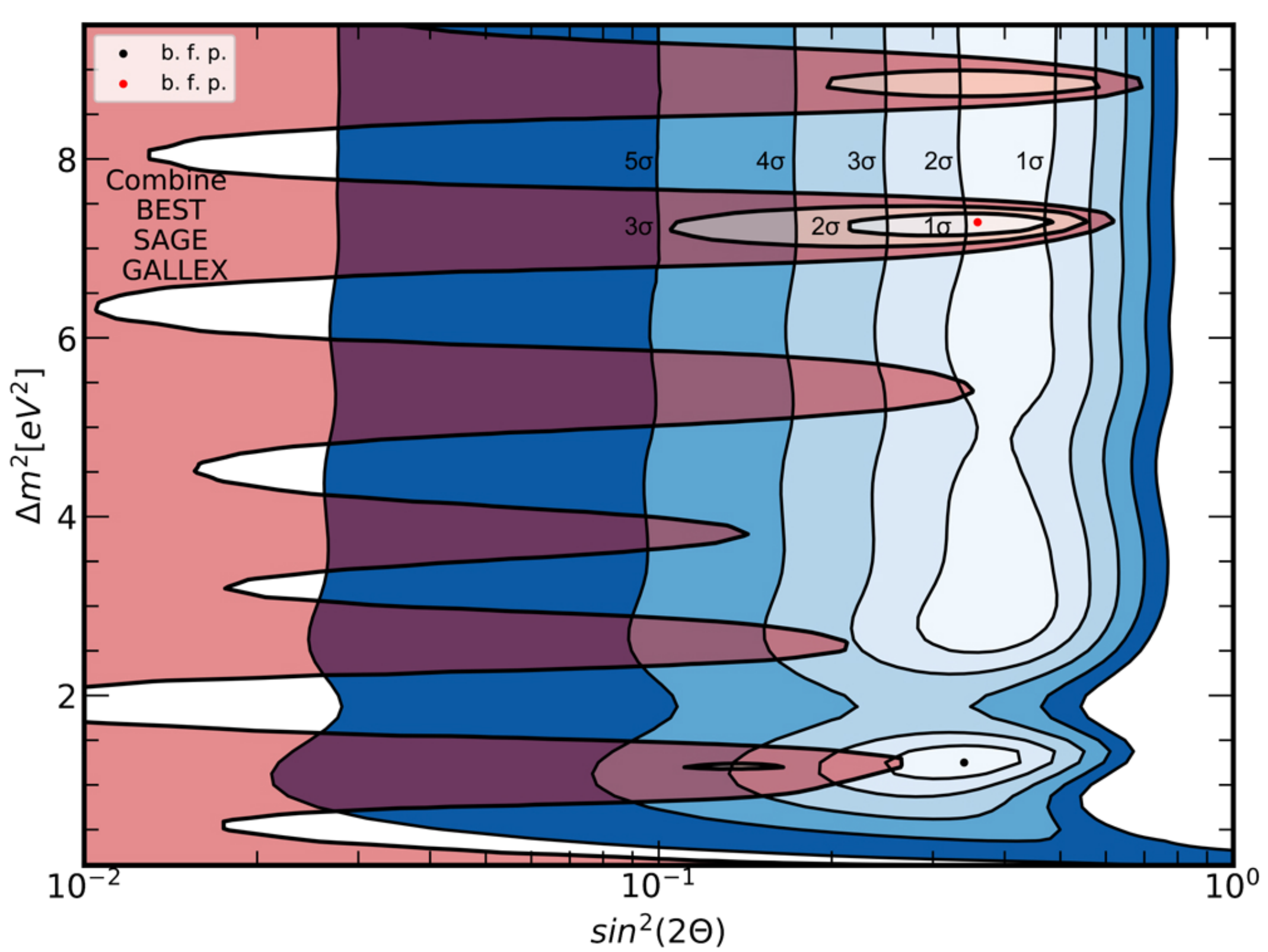}}
  \caption{The regions (in shades of blue)
    {\it favored} by all the gallium experiments\,\cite{Barinov:2021asz} for mixing with
  electron neutrinos. Assuming the same mixing with electron
  antineutrinos there are also regions (in shades of pink)
  {\it favored} by NEUTRINO-4\,\cite{Serebrov:2020kmd}.}
\label{fig:gal}
\end{figure}

BEST exploited an artificial chromium source 
simultaneously irradiating two volumes filled with gallium, which
allowed one to measure the neutrino flux at two different distances.  
The BEST experiment 
has published results\,\cite{Barinov:2021asz}, which confirm
the gallium anomaly and are consistent with its explanation as electron neutrino
oscillations into the sterile components.
In the BEST experiment the artificial source was very compact, placed
in the center of the inner spherical volume, which was placed in the center of
the outer cylindrical volume. 
The neutrino fluxes were
measured as $R_{in}=0.791\pm0.05$ and $R_{out}=0.766\pm 0.05$ from
what are expected, which implies more than 5\,$\sigma$ evidence for
the disappearance of electron neutrinos. Within the sterile neutrino
hypothesis this disappearance is explained as oscillations into
sterile neutrino states, so that the electron neutrino flux at a given
distance $r$ from the point-like source is suppressed by the survival
probability
\begin{equation*}
P\l E_\nu,r\r=1-\sin^22\theta\,\sin^2\!\!\l \!\! 1.27\times 
\frac{\Delta m^2[\eV^2] 
\,r[\text{m}]}{E_\nu[\MeV]}\!\r\!.
\end{equation*}
 With the detector size of about a meter and
the neutrino energy of about 750\,keV (the dominant line of the chromium
source) it allows for testing the oscillations with $\Delta m^2\sim
1$\,eV$^2$.   
The geometry details are presented in Ref.\,\cite{Barinov:2021asz}, 
and the sterile neutrino parameters (mass and mixing angle) are
defined from the likelihood function given by eqs.\,(3)-(5) of
Ref.\,\cite{Barinov:2021asz}. Then repeating the joint analysis for the
all gallium experiments one obtains the favorable region in the model
parameter space outlined in Fig.\,\ref{fig:gal};
the 1-,2-
and 3-$\sigma$ contours coincide with those in Fig.3 of
Ref.\,\cite{Barinov:2021asz}. We observe that {\it within the sterile
neutrino hypothesis} the gallium experiments provides higher
than 5\,$\sigma$ evidence for the presence of sterile
neutrinos. Actually, one can find that even the single BEST result exceeds
5\,$\sigma$ level.  

%{\bf 3.}
\section{Joint analysis}
The results obtained by the BEST experiment favor rather
large mixing angle and either the region of $\Delta m^2\simeq
1$\,eV$^2$ or noticeably larger masses $\Delta m^2\gtrsim 
3$\,eV$^2$. The likelihood contours on the plot of  
Fig.\,\ref{fig:gal} show that in the region of large masses
the BEST has no sensitivity to the sterile neutrino
masses, which correspond to the oscillation lengths shorter
than the typical meter-scale size of the BEST detector volumes. Remarkably, parts 
of this region have been favored by the anomalous results (2.8\,$\sigma$)
of one of the reactor neutrino experiments,
NEUTRINO-4\,\cite{NEUTRINO-4:2018huq,Serebrov:2020kmd}. We plot these regions in
Fig.\,\ref{fig:gal} for comparison. Note that the Neutrino-4 likelihood exhibits a
local minimum in the region of best fit point (b.f.p.) of the joint gallium anomaly,
$\Delta m^2\simeq 1$\,eV$^2$. There are also minima at large masses,
and the global minimum at $\Delta m^2\simeq$7\,eV$^2$, where
1-$\sigma$ contours of NEUTRINO-4 and gallium experiments overlap. 

\begin{figure}[!htb]
  \centerline{
\includegraphics[width=\columnwidth]{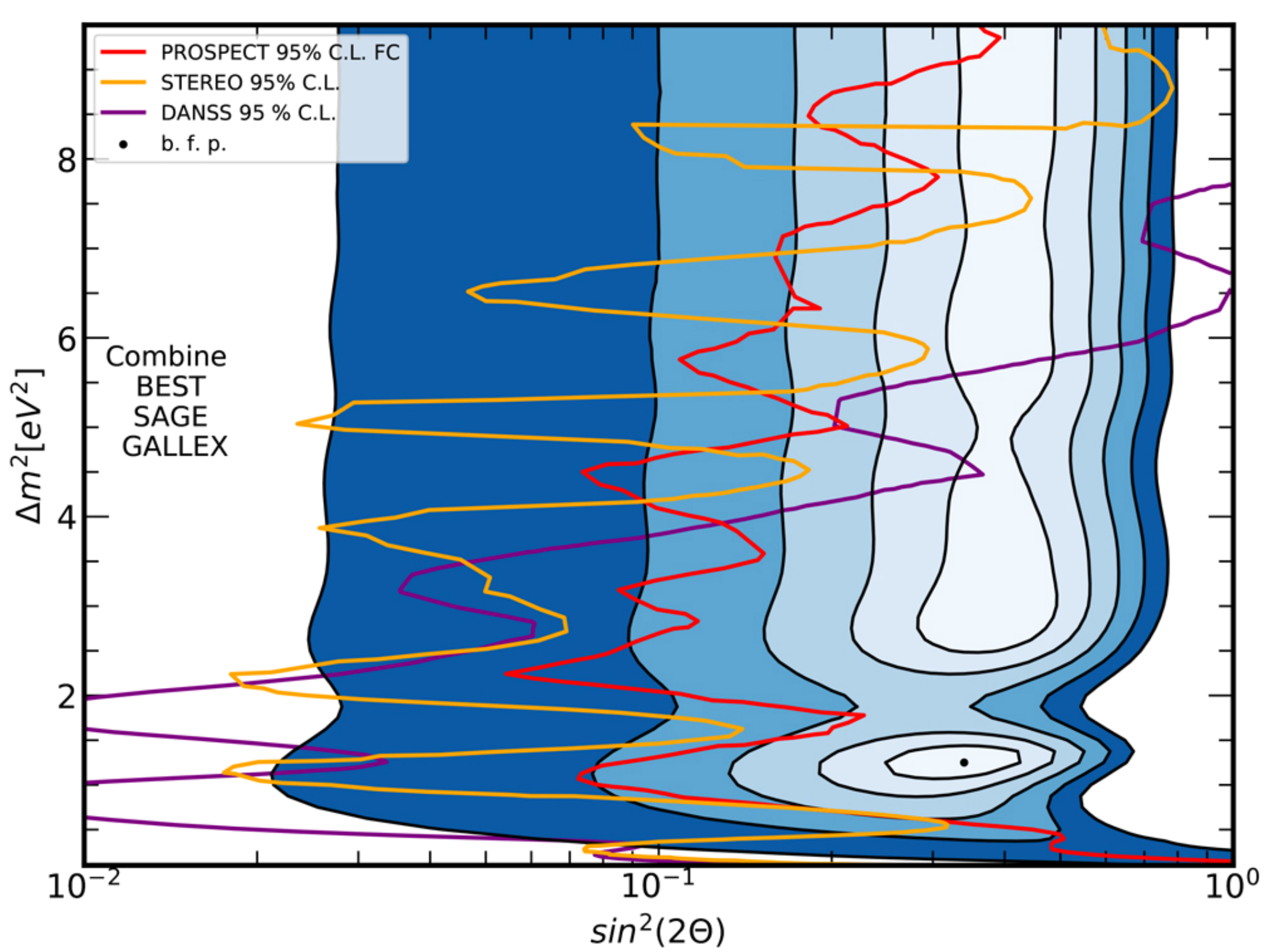}}
\caption{The regions {\it favored} by all the gallium
  experiments\,\cite{Barinov:2021asz} (the same as in
  Fig.\,\ref{fig:gal}) for mixing with
  electron neutrinos and limits from searches for sterile neutrinos at
  reactor antineutrino experiments STEREO\,\cite{STEREO:2019ztb},
  PROSPECT\,\cite{PROSPECT:2020sxr} and DANSS\,\cite{Danilov:2020ucs}. The regions of large mixing are
  {\it excluded} by each experiment at 95\% C.L.}
\label{fig:reactor}
\end{figure}

The region of masses $\Delta m^2\simeq 1$\,eV$^2$ has been thoroughly
explored by searches for sterile neutrino in the reactor antineutrino experiments, see
Fig.\,\ref{fig:reactor},  
the strongest constraint
obtained by the DANSS
experiment\,\cite{DANSS:2018fnn,Danilov:2020ucs}. Note, that while at
small masses the favored by gallium anomaly region is heavily
constrained by these searches, at large masses, $\Delta
m^2>5$\,eV$^2$, there are 2-$\sigma$ and even 1-$\sigma$ patches favored by
gallium experiment and consistent with all these bounds.  

\begin{figure}[!htb]
\centerline{
  \includegraphics[width=\columnwidth]{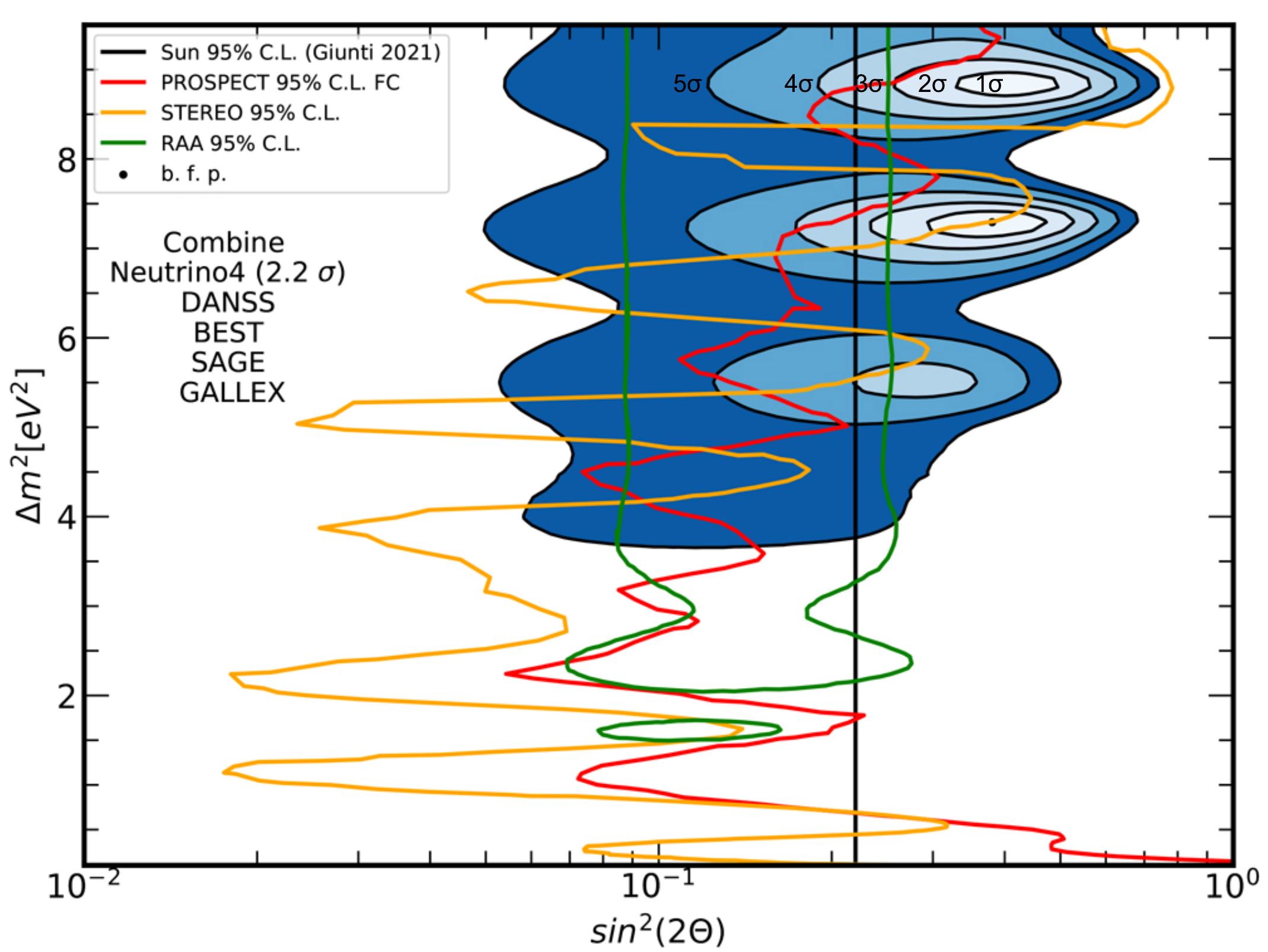}}
\caption{The regions (in shades of blue) {\it favored} by the joint analysis of the gallium
  experiments, DANSS\,\cite{Danilov:2020ucs} and
  NEUTRINO-4\,\cite{Serebrov:2020kmd}. 
  There are also regions {\it excluded} at 95\% C.L. from
  sterile neutrino searches at reactor antineutrino experiments STEREO\,\cite{STEREO:2019ztb},
  PROSPECT\,\cite{PROSPECT:2020sxr}. The regions outlined by the green
  line is {\it favored} at 95\% C.L. by the reactor antineutrino anomaly
  (RAA)\,\cite{Abazajian:2012ys}. The region to the right of the
  black vertical line is {\it excluded} at 95\% C.L. from observations of
  solar neutrinos\,\cite{Giunti:2021iti}.}
\label{fig:joint}
\end{figure}

Encouraged by the overlap presented on the plot of
Fig.\,\ref{fig:gal}, we perform a joint analysis of the
all gallium experiments, NEUTRINO-4 and DANSS, which yields the
likelihood contours 
depicted in Fig.\,\ref{fig:joint}.
We utilize here the original DANSS likelihood\,\footnote{We thank M.\ Danilov and
N.\ Skrobova for sharing the
DANSS likelihood distribution.} and the NEUTRINO-4 likelihood\,\footnote{We thank A.\ Serebrov and
R.\ Samoilov for sharing the
Neutrino-4 likelihood distribution.}
somewhat corrected as advertised
in Ref.\,\cite{Giunti:2021iti} to be conservative (the NEUTRINO-4 anomaly becomes then at
2.2\,$\sigma$ level). The joint anomaly inferred from this analysis is at about 5.7\,$\sigma$, and
it favors the region of large sterile neutrino masses, $\Delta m^2>5$\,eV$^2$. The
significance level is dominated by the BEST anomalous result, while
the positions of the local minima of the joint likelihood at
particular values of the sterile neutrino mass are mostly 
determined by those of the NEUTRINO-4 experiment. 

We also plot in Fig.\,\ref{fig:joint} limits from sterile neutrino
searches at STEREO and PROSPECT experiments (the regions of large
mixing are {\it excluded} at 95\% C.L.), and the green contour that
outlines the region {\it favored} at 95\% C.L. by the reactor
antineutrino anomaly (RAA), see e.g.\,\cite{Abazajian:2012ys}.  One
observes, that the STEREO limits are consistent with the joint anomaly
presented in Fig.\,\ref{fig:joint}, while the limits from PROSPECT
disfavor 2-$\sigma$ regions except a small part at $\Delta m^2\approx
9$\,eV$^2$. The reactor antineutrino anomaly region also has a small
spot with 2-$\sigma$ regions of the joint anomaly.

The analysis above shows that the BEST anomalous result, if explained
within the hypothesis of a single sterile neutrino (dubbed $3+1$ scheme) points at the region of large
mixing and masses, which will be explored at the next operation stages
of the  upgraded reactor
neutrino experiments and KATRIN experiment on tritium $\beta$-decay,
see Ref.\,\cite{KATRIN:2020dpx}. The position of the best fit point
and the likelihood contours 
may be refined with a joint statistical analysis including the
likelihoods of other reactor antineutrino experiments relevant for the
task. 

Finally, recall that, apart from the direct limits, the typical $3+1$
neutrino model is strongly constrained from astrophysics and excluded by
cosmological observations within simple extensions of the Standard
Cosmological Model. Indeed, measurements of the solar neutrino flux
exclude the models with large mixing\,\cite{Giunti:2021iti},
the black vertical line in Fig.\,\ref{fig:joint} refers to the
corresponding 95\% C.L. bound. Likewise, the large sterile-active neutrino mixing produces sterile
neutrinos in the plasma of the early Universe in the amount forbidden
from analysis of present cosmological data\,\cite{Planck:2018vyg}. It
is tempting to suggest a modification of the simple $3+1$ scheme,
which would circumvent the both indirect constraints.

\section{Conclusions}
%{\bf 4.}
To summarize, we investigate how the recent results of BEST
experiment can affect the hypothesis of '3+1' scheme in neutrino
sector. Note in passing that the explanation of the observed lack of neutrino
events as oscillations to the sterile neutrino is not the only
possibility involving new physics. Moreover, an overall error in
neutrino cross section on gallium or some issues in extraction
efficiency cannot be ruled out, though all the aspects of experimental
procedure have been verified and many have been 
double-checked\,\cite{Barinov:2021asz}.

We thank M.\ Danilov, V.\ Rubakov and S.\ Troitsky for
stimulating questions and remarks about the results of the BEST
experiment. The work was supported by the the RSF grant
17-12-01547.

%%%%%%%%%%%%%%%%%%%%%%%%%%%%%%%%%%%%%%%%%%%%%%%%%%%%%%%%%%%%%%%%%%%

%%%%%%%%%%%%%%%%%%%%%%%%%%%%%%%%%%%%%%%%%%%%%%%%%%%%%%%%%%%%%%%%%%%
\bibliographystyle{apsrev4-1}
\bibliography{C}

\end{document}